# Role of the chiral spin configuration in field-free spin-orbit torque-induced magnetization switching by a locally injected spin current

Suhyeok An[1], Hyeong-Joo Seo[1], Eunchong Baek[1], Soobeom Lee[2], Chun-Yeol You[1,*]

[1] Department of Physics and Chemistry, DGIST, Daegu 42988, South Korea

[2] Emerging Materials Science Research Center, DGIST, Daegu 42988, South Korea

[*] Correspondence to cyyou@dgist.ac.kr

For deterministic magnetization switching by spin-orbit torque (SOT) in a perpendicular magnetic anisotropy system, an additional in-plane direction magnetic field is essential for deterministic switching by breaking the magnetization symmetry. Realizing chirality in a magnetic ordering system can be one approach for achieving asymmetry in the lateral direction for field-free magnetization switching. However, systematic analysis of the influence of the chiral spin system on deterministic switching is still scarce. In this report, the achievement of field-free SOT-induced magnetization switching by using a chiral spin configuration is investigated by experiments and micromagnetic simulations. We designed a system in which only part of the ferromagnetic layer overlaps with the heavy metal layer in the Pt/Co/MgO structure. Therefore, a spin current is exerted only on a local area of the ferromagnetic layer, which results in a Néel-type chiral spin configuration. The induced chiral spin configuration can be stabilized (or destabilized) depending on the sign of the interfacial Dzyaloshinskii-Moriya interaction of the system. The stabilized spin configuration plays a crucial role in the deterministic switching in zero field. We expect our findings to widen the perspective on chirality-based all-electrical SOT device fabrication.

## I. Introduction

Since the discovery of magnetization control by the spin-orbit coupling-induced spin transfer torque effect in heavy metal (HM)/ferromagnetic metal (FM) layers[1,2], i.e., the so-called spin-orbit torque (SOT), SOT-induced magnetization switching has attracted great interest because of its superior benefits of fast switching speed and efficient power consumption[3-9]. The SOT-induced magnetization switching mainly originates from the spin Hall effect (SHE) of the HM layer and/or the Rashba effect of the interface between the HM and FM, so easier/faster switching is achieved by a larger spin current. Here, the spin Hall angle (SHA) indicates the degree of conversion of the electric charge current to the spin current by the SHE. Therefore, SHAs have been heavily studied in various HMs[10-13], HM-normal metal (NM) alloys[14-22], and HM/NM multilayers[23,24].

However, to achieve deterministic SOT-induced magnetization switching in a perpendicular magnetic anisotropy (PMA) system, symmetry breaking of the magnetization in the in-plane direction is necessary. Applying an additional external magnetic field parallel to the charge current is a commonly used method for breaking the in-plane symmetry of the magnetization. However, magnetization switching using only a current is essential for practical device applications, and therefore, many studies on field-free SOT switching methods are under investigation as follows.

There are several categories in field-free SOT switching. One category is additional lateral symmetry breaking at the nanoscale. Yu *et al.* [25] reported that field-free SOT switching was achieved in a Ta/CoFeB/TaO$_x$ structure. They used wedged top Ta capping layer oxidation to introduce lateral asymmetry caused by the PMA gradient in the in-plane direction. A similar result was reported for the wedged FM layer in the Ta/CoFeB/MgO structure[26]. Local oxidation of the FM layer creates a magnetic anisotropy gradient system[27]. The oxidized part of the nanostructure loses its PMA, while the other unoxidized part retains its PMA. Combining the Dzyaloshinskii-Moriya interaction (DMI) with such nonuniform magnetic anisotropy in the nanostructure, a chiral spin configuration is formed at the boundary between the PMA and in-plane anisotropy regions, and field-free SOT switching is achieved by the chiral spin configuration. In addition, a system with internal asymmetry can also achieve magnetic field-free switching, such as a system with a spin current gradient[28] and tilted PMA[29]. Another category is methods using additional layer structures: antiferromagnets[30,31], FM trilayers[32], interlayer exchange coupling[33], and competing spin current layers[34].

In most cases, field-free SOT switching methods require special nanoscale engineering, such as wedge-type deposition, local oxidation, and additional layers. All these approaches require additional costs in the actual device process. Therefore, magnetic field-free switching only in the well-known standard HM/FM structure is expected to lead to an improvement in the efficiency of the fabrication of SOT-based devices because SOT phenomena that have been heavily studied previously can be applied.

Here, we report that SOT-induced field-free switching is achieved with the chiral spin configuration created by a locally injected spin current with the DMI in a Pt/Co/MgO structure with PMA. To fabricate the system of a locally injected spin current, we designed a slab structure of an FM whose one axis is longer than the width of an HM, as shown in Fig. 1(a). The spin current from the HM is only injected into a local part of the FM layer (overlapped area indicated as a dark green colored region in Fig. 1(a)). The magnetization of the SOT-influenced area aligns with the injected spin current polarization ($y$-axis), while the noninjected (uninfluenced) area magnetization remains in its initial state ($+z$-axis). As a result, a Néel-type chiral spin configuration is formed in the influenced and uninfluenced areas. The interplay between the chiral spin configuration created by the locally injected spin current and the proper sign of interfacial DMI energy density ($D_{int}$) enables SOT-based field-free switching, which is the key idea of the present work.

## II. Results and Discussion
### A. Sample preparation

We deposited a thin film sample of a Ta(3)/Pt(5)/Co(1.2)/MgO(2)/Ta(2) stack on a single-sided polished silicon oxide substrate. Here, the unit of the numbers in parentheses is nm. We used a magnetron sputtering system for sample deposition, DC power sources for metals such as Ta, Pt, and Co, and an AC source for MgO. Here, the upper and lower Ta layers are a buffer for PMA and a capping layer for sample protection, respectively. After the deposition process was completed, the HM line and FM slab fabrication processes proceeded. The processes were performed through two e-beam lithography patterning and Ar ion beam etching processes. The first etching process was a full-layer etching process to form the Hall bar geometry, and the second etching process was a selective etching process to make the slab shape of the FM layer. Therefore, in the second etching, the process was conducted until only the FM layer was etched, as seen in Fig. 1(b), which was obtained by atomic force microscopy. Since the FM layer is etched except for the slab structure, all electrical signals are observed through the HM layer. The width of the HM line of the processed sample is 2 μm, and the size of the FM slab is 1.2 μm width × 10 μm length. The overlapping part of the HM line and the FM slab is only a 1.2 μm width × 1.5 μm length area, indicated by the red dashed box in Fig. 1(b). The SOT effect on the FM layer appears only in this influenced area. After the etching process was finished, Ta(10)/Cu(75) electrodes were deposited, and the anomalous Hall effect (AHE) was measured to confirm whether PMA was formed. A 50 µA DC was injected along the *x*-axis direction, and the magnetic field was swept in the *z*-direction (out-of-plane). As shown in Fig. 1(c), the sample showed a clear hysteresis loop, indicating that the sample has sufficient PMA. The observed AHE voltage ($\Delta V_{H,AHE}$) is approximately 18.5 µV.

**B. SOT-induced magnetization switching**

The SOT-induced magnetization switching was measured in several fixed external in-plane magnetic fields, as shown in Fig. 2. Here, the direction of the external magnetic field and the pulse current was along the ±*x*-axis direction. An initialization process was performed immediately before the measurement using ±*z*-axis direction external magnetic fields larger than the coercivity to align the magnetization in the ±*z* (up or down) direction. After initialization, a field $B_x$ was applied to observe the tendency for external field-dependent deterministic switching. The current pulse was injected with amplitude ($I_P$) from –8.5 mA (8.5 mA) to 8.5 mA (–8.5 mA) at a 100 µA step for the initial state of up (down) magnetization. Here, the pulse width was 1 ms. To measure the magnetization state after SOT, a small reading current of 50 µA DC was applied to measure $V_H$ between each current pulse. Notably, the measured result showed clear magnetization switching not only in the absence of an external magnetic field (field-free) but also under a negative (-*x* direction) external magnetic field, where magnetization switching normally does not occur. Here, the solid curve and dashed curve indicate measured data after up and down direction initialization.

To analyze the SOT switching tendency, the critical switching current ($I_{P,Crit}$) was extracted from the results in Fig. 2 according to $B_x$. From the macrospin model, if $B_x$ is sufficiently small compared to the effective PMA field, then the SOT-induced magnetization switching tends to linearly decrease with $B_x$[4], which is well shown in Fig. 3(a). Notably, the switching occurs even when the external magnetic field is applied in the negative direction beyond zero field. This means that approximately –10 mT is required to compensate for an effective internal effect that is formed by the locally injected spin current. The degree of magnetization switching by SOT can be detected by comparing the $V_H$ values before and after measurements, represented as $\Delta V_{H,\text{switching}}$ in the inset of Fig. 2. $A_{switching}$ is the portion that is determined by the ratio of the up and down domains inside the SOT-influenced area and can be calculated following $A_{switching} = \Delta V_{H,\text{switching}}/\Delta V_{H,AHE} \times 100(\%)$. Therefore, for $A_{switching}$, a value closer to 100% means that most of the magnetization in the SOT-influenced area is switched from the initial state. We depict $A_{switching}$ as a function of $B_x$ in Fig. 3(b) for the up (down) to down (up) case. The plot is clearly shifted to a nonzero negative $B_x$, which is clear evidence of breaking of the in-plane directional symmetry. In the up (down) to down (up) switching, 85.2% (61.3%) of the area is switched in zero field. Note that partial switching may not be a problem in real devices, in which we may design the reading signal to be only from the switched area. This is because the switched area is well defined by the spin current injected area. An adequate $A_{switching}$ value in zero field means that the internal effect caused by the locally injected spin current is sufficiently large to cause magnetization switching without an external magnetic field.

**C. Micromagnetic simulations under various conditions**

To understand the origin of the internal field caused by the local spin current, a micromagnetic simulation was conducted by Mumax$^3$[35]. The grid size was set as 120 nm width × 300 nm length ($L$) × 1 nm thickness, and the cell size was 1 nm. The material parameters used were as follows: $M_s$ = 1.0 MA/m; $A_{ex}$ = 20 pJ/m; $\alpha$ = 0.02; $K_U$ = 1.0 MJ/m$^3$; and $L$ = 300 nm. Here, $M_s$, $A_{ex}$, $\alpha$, $K_U$, and $L$ are the saturation magnetization, exchange stiffness, damping constant, first-order uniaxial anisotropy energy, and length of the FM slab, respectively. To implement the PMA state, the easy axis was set to the $z$-axis. To simulate the magnetization switching behavior of the PMA influenced by the SHE, we set the pure spin current to flow in the $z$-direction with a polarization along the +$y$-direction. Here, the spin current was injected with a duration of 1.0 ns, and the rising and falling times were set as 0.5 ns with a hyperbolic tangent form. The peak amplitude of the pure spin current pulse ($J_{Peak}$) was 1.5 × 10$^{12}$ A/m$^2$. In all our simulations, the spin current affected part of the FM layer was set to a 120 nm width × 100 nm length area, as indicated by the blue dashed box in Fig. 4(a). The plotted $m_y$ and $m_z$ are considered only in this influenced area. The simulations in Fig. 4 were performed without an external field. When the spin current is injected, the magnetization in the spin current injection area is aligned in the spin polarization direction (+$y$), as seen in the 0.7 ns result in Fig. 4(a) and Fig. 4(b). Thus, the FM slab has a Néel-type chiral spin configuration along the $y$-axis. Then, when the current is turned off, the magnetization in the spin current injected area switches, resulting in a Néel-type domain wall, as in the 2.2 ns result in Fig. 4(b). This series of processes is caused when the spin configuration has an appropriate Néel-type chirality. Therefore, the effect of $D_{int}$ is expected to have a great influence on the magnetization switching.

For more details about the relation between the interfacial DMI and the deterministic switching, we performed micromagnetic simulations in 2 cases. The first concerns the possibility of deterministic switching according to $B_x$ and $D_{int}$. In this case, the grid size was 120 nm width × 600 nm length ($L$) × 1 nm thickness. The material parameters used were as follows: $M_s$ = 1.0 MA/m;

$A_{ex} = 20$ pJ/m; $\alpha = 0.02$; $K_U = 0.7$ MJ/m$^3$; and $J_{Peak} = 8.55 \times 10^{11}$ A/m$^2$. Fig. 5(a) shows the time-dependent $m_z$ according to $B_x$ with fixed $D_{int} = 0.7$ mJ/m$^2$. Here, the black dotted line corresponds to the normalized spin current density during the simulation. We omit the positive field results since the field-assisted SOT switching is redundant. We would like to emphasize that deterministic switching is achieved not only at zero field but also with a finite negative field (> –6 mT), where the negative field suppresses SOT-induced switching from the up to down state. These results are similar to the experimental results (see Fig. 2): a finite negative field is required to compensate for the finite effective internal field that promotes field-free SOT switching. In addition, a similar tendency is also observed in the case of different $D_{int}$, as seen in Fig. 5(b) with fixed $B_x = –5$ mT. Even though $B_x$ is negative, switching occurs when $D_{int}$ ($\geq 0.7$ mJ/m$^2$) is large enough, and switching fails when $D_{int}$ ($\leq 0.6$ mJ/m$^2$) is small. This indicates that the switched and nonswitched states are governed by the interplay of $B_x$ and $D_{int}$. For a wider perspective, we studied the switching condition by mapping the switching according to $B_x$ and $D_{int}$, as seen in Fig. 5(c). Each range of $B_x$ and $D_{int}$ for the mapping diagram is as follows: ($B_{x,min} = 0.0$ mT, $B_{x,gap} = –0.2$ mT, $B_{x,max} = –10.0$ mT) and ($D_{int,min} = 0.40$ mJ/m$^2$, $D_{int,gap} = 0.01$ mJ/m$^2$, $D_{int,max} = 1.00$ mJ/m$^2$). The typical $D_{int}$ values for similar samples from our group are ~0.8 mJ/m$^2$[36, 37]. Note that we are mapping only the negative $B_x$ and positive $D_{int}$ results because the positive $B_x$ case always shows switched results (a positive $B_x$ can be used for field-assisted deterministic switching from the up to down state). Additionally, a negative $D_{int}$ gives us the same results when the spin current direction is reversed. See Fig. S1(a) in the Supplementary Material for the current direction-dependent simulation results. All $m_z$ values are extracted at 2 ns from the time-dependent results. Because complex domain wall dynamics appear over time, obtaining a clear boundary curve between the switched and nonswitched states according to $B_x$ and $D_{int}$ is difficult. Nevertheless, Fig. 5(c) shows that the switched and nonswitched states are divided based on a specific region (green region). Here, because the torque induced by $B_x$ affects the magnetization such that it tends toward the nonswitched state, the result from Fig. 5(c) implies that the additional effect of the local spin current better overcomes the negative $B_x$-induced torque as $D_{int}$ increases.

More simulations were carried out to confirm the role of the sign of $D_{int}$ and the stray field from the uninfluenced area of the FM slab. We changed the sign of $D_{int}$ and $L$ of the system. As seen in Fig. 5(d), the tendency for magnetization switching and nonswitching depends on the $D_{int}$ sign as we expected. This trend suggests that a negative $D_{int}$ prefers the opposite chirality compared with the chirality induced by the locally injected SOT. As a result, the spin dynamics show much more complex behavior and are not switched, as shown by the red line in Fig. 5(d). Since the FM layer takes the form of a slab, there is a nonzero stray field from the uninfluenced area of the FM. Because the strength of the stray field should depend on $L$, the time-dependent $m_z$ is plotted at various $L$ values in Fig. 5(e). The switching behavior is insensitive to $L$ when $L \geq 110$ nm. The results tell us two things. First, the role of the stray field in the observed field-free switching is not important. Second, once the chiral spin configuration is formed by the locally injected spin current, deterministic field-free switching occurs despite $L$. However, when all of the area of the FM is influenced by SOT, field-free switching fails. For a wider perspective, each switching result for each $D_{int}$ sign and various values of $L$ is shown in Fig. 5(f). At an $L$ of 100 nm or less, the spin current is applied to all of the FM layer, resulting in a uniform spin configuration in the whole system. However, at $L$ larger than 100 nm, the Néel-type chiral spin configuration is formed by the locally injected spin current, resulting in a difference in switching according to the $D_{int}$ sign. In addition, since the energy of the chiral structure is different according to the sign of the system DMI[38], switching also does not occur if the spin current polarization direction is reversed. See Fig. S1(b) in the Supplementary Material for the possibility of switching at positive $D_{int}$. The $D_{int}$- and $L$-dependent switching results can be interpreted to indicate that the interplay between the SOT-induced chiral spin configuration and the DMI of the sample is a key component of deterministic field-free switching.

## III. Conclusions

In this work, we experimentally observed field-free SOT-induced magnetization switching by a locally injected spin current combined with the DMI using a slab-shaped FM layer. To reveal the physical origin of our findings, we carried out micromagnetic simulations. We found that the locally injected spin current with the proper sign of $D_{int}$ generates a stable Néel-type chiral spin configuration. When the current is turned off, the stabilized chiral spin configuration leads to deterministic field-free switching. Our results are applicable to the data input part of domain wall-based logic devices, such as spin torque majority gates[39,40], and are expected to provide a wider perspective on chirality-based all-electrical SOT device applications.

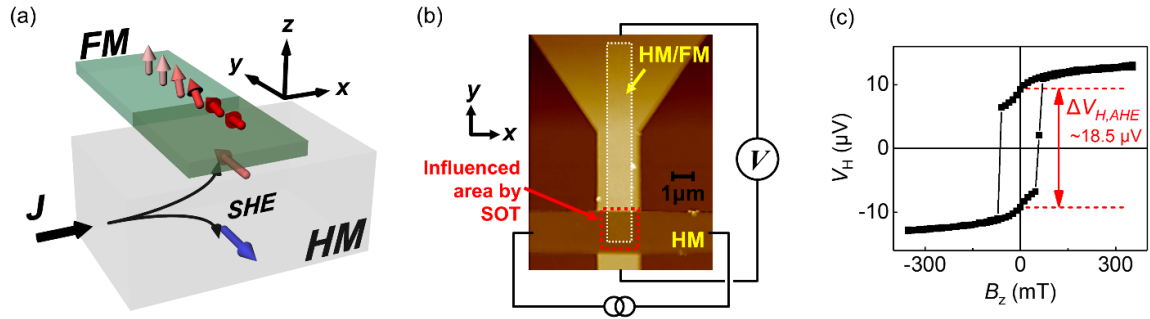

**Fig. 1**. (a) Schematic diagram of a locally injected spin current in a slab-shaped FM layer. Only the area overlapping with the HM is influenced by SOT (dark green-colored area of the FM layer), resulting in alignment of the magnetization along the ±y-axis, and a chiral spin configuration is formed at the edge of the influence area. (b) Fabrication result of a device measured by atomic force microscopy. The area influenced by the SOT is shown with a red dotted box. A schematic of the circuit is also shown. (c) AHE hysteresis loop measured under an external magnetic field with the direction of the magnetic easy axis (z-axis). Here, $\Delta V_{H,AHE}$ is the difference in the AHE voltage when the external magnetic field is zero.

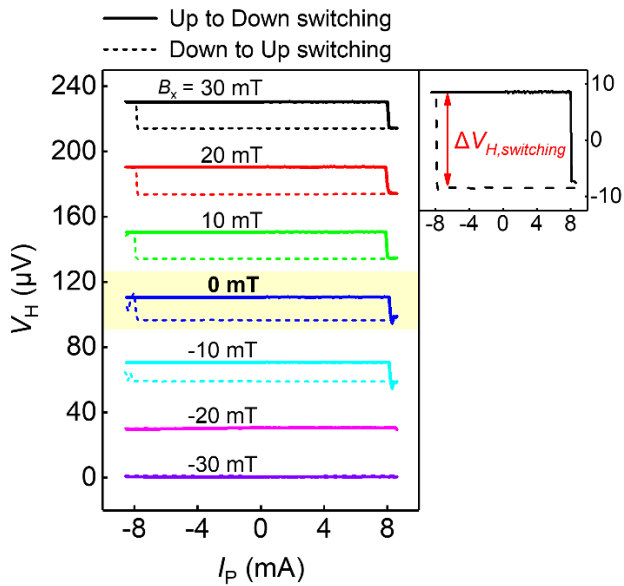

**Fig. 2**. SOT-induced magnetization switching results for various external magnetic fields according to the magnetization initialization direction. The solid line indicates up direction initialization, and the dashed line indicates down direction initialization. Here, the inset shows an enlarged result measured under an external magnetic field of 30 mT. $\Delta V_{H,switching}$ is determined by calculating the

$V_H$ difference before and after SOT-induced magnetization switching, |$V_H$ (before switching) – $V_H$ (after switching)|.

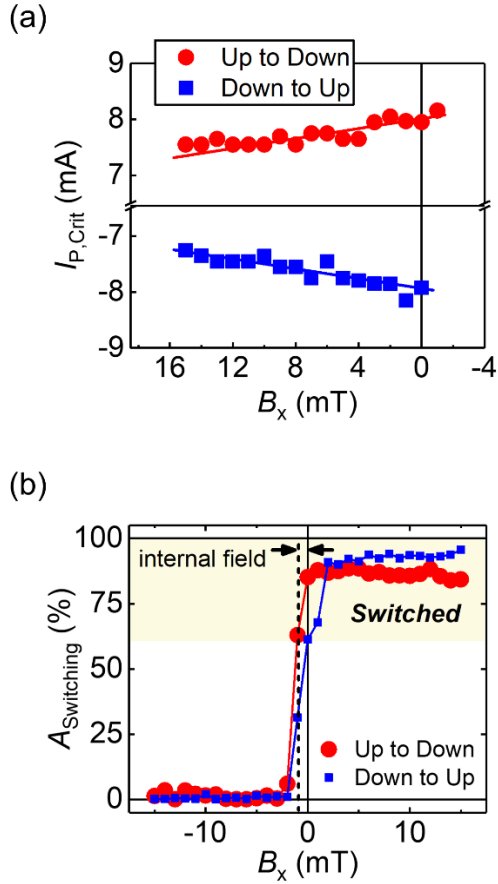

**Fig. 3.** (a) Switching critical current as a function of the external magnetic field for each switching direction. Here, each straight line is the result of linear fitting. (b) Proportion of the down (up) domains formed by SOT-induced magnetization switching in up-to-down switching (down-to-up switching), calculated with $A_{switching} = \Delta V_{H,switching}/\Delta V_{H,AHE} \times 100\%$. Here, the indicated "switched" (yellow) area means that 60% or more of the magnetization is switched.

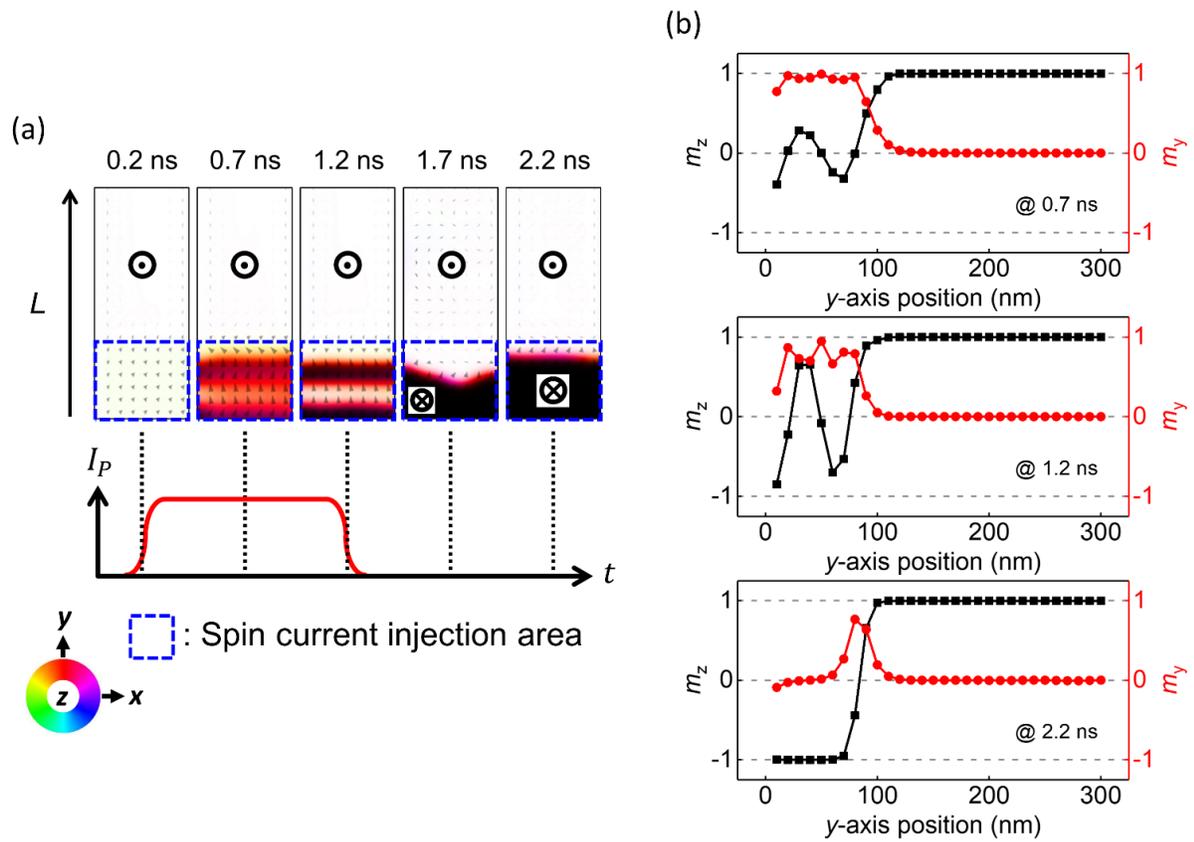

**Fig. 4.** (a) Time-dependent magnetization colormap. Here, each color indicates the direction of magnetization, as shown in the color bar on the right side. White (black) is up (down) direction magnetization. The below graph is a schematic diagram of the spin current pulse, and the spin current is injected only into the area indicated by the blue dashed rectangle. (b) $m_y$ and $m_z$ values according to the y-axis position of the FM slab at 0.7 ns, 1.2 ns and 2.2 ns.

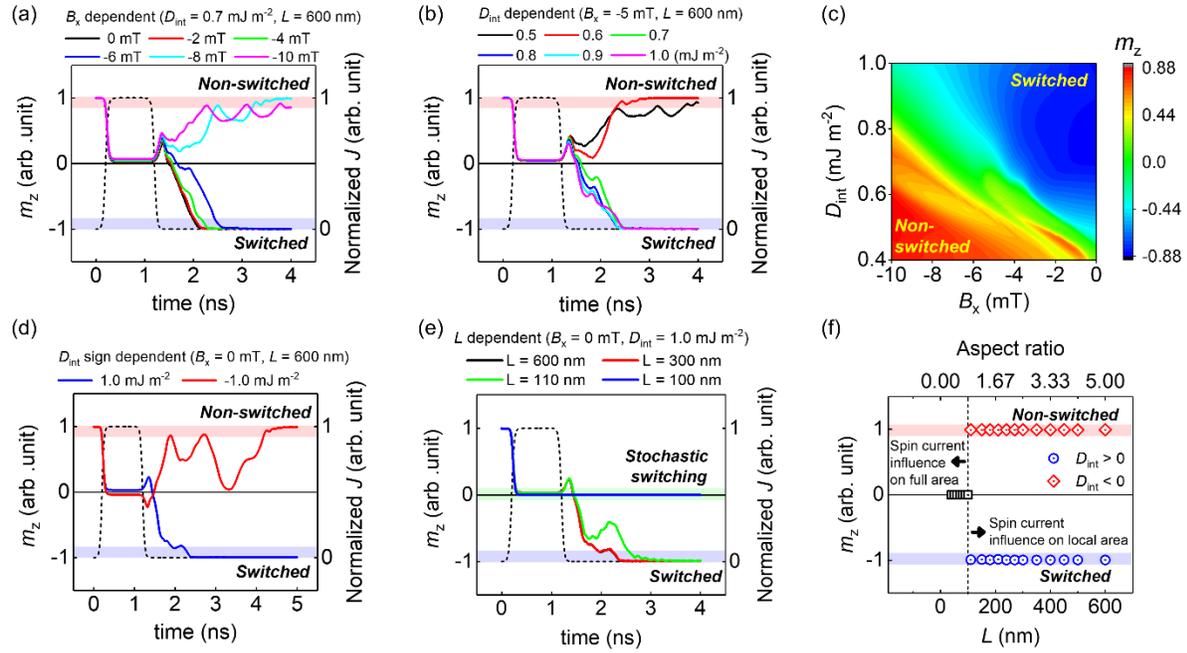

**Fig. 5.** Micromagnetics results of SOT-induced magnetization switching according to interfacial DMI energy density ($D_{int}$), external magnetic field ($B_x$) and length of the FM slab ($L$). The initial state is a fully saturated up state ($m_z = +1$). Time-dependent $m_z$ result at (a) fixed $D_{int}$ ($D_{int} = 0.7$ mJ/m$^2$) for various $B_x$ and (b) fixed $B_x$ ($B_x = -5$ mT) for various $D_{int}$. Here, the black dotted line represents the pulse amplitude over time. (c) SOT-induced magnetization switching mapping result according to $D_{int}$ (> 0.4 mJ/m$^2$) and $B_x$ (≤ 0 mT). (d) $D_{int}$ sign- and (e) $L$-dependent $m_z$ results at zero field. (f) Possibility of deterministic switching according to the aspect ratio of the FM slab and $D_{int}$ sign.

**Acknowledgements**

This work is supported by the National Research Foundation of Korea (Grant Nos. 2020M3F3A2A02082437, 2021R1A2C2007672, 2021M3F3A2A01037525).